\documentclass[aps,prb,preprint,groupedaddress,showpacs]{revtex4}
\usepackage{graphicx}
\usepackage{amssymb}
\usepackage{amsmath}
\usepackage{latexsym}
\usepackage{ulem}
\usepackage{color}
\usepackage{dcolumn}
\usepackage{subfigure}

\begin{document}

\title{Reconsidering the possibility of room temperature ferromagnetism in Mn doped Zirconium oxide}

\author{Akash Chakraborty}
\email{akash.chakraborty@physik.uni-regensburg.de}
\affiliation{
Institut f\"ur Theoretische Festk\"orperphysik, Karlsruhe Institute of Technology, 76128 Karlsruhe, Germany
}
\affiliation{
School of Engineering and Science, Jacobs University Bremen, Campus Ring 1, 28759 Bremen, Germany
}
\author{Georges Bouzerar}
\email{georges.bouzerar@univ-lyon1.fr}
\affiliation{
Institut Lumi\`ere Mati\`ere, Universit\'e Lyon 1-CNRS, F-69622 Villeurbanne Cedex, France
}

\date{\today}

\clearpage

\begin{abstract}
The possibility to induce long range ferromagnetic order by doping oxides with transition metal ions has become 
a very exciting challenge in the last decade. Theoretically, it has been claimed that Mn doped ZrO$_2$ could be a very promising spintronic candidate and that 
high critical temperatures could be already achieved even for a low Mn concentration. Some experiments have reported room temperature ferromagnetism (RT-FM) whilst 
some others only paramagnetism. When observed, the nature of RT-FM appears to be controversial and not clearly understood. In this study, we propose to  clarify and 
shed light on some of theses existing issues. A detailed study of the critical temperatures and low energy magnetic excitations in Mn doped  ZrO$_2$ is performed. We 
show that the Curie  temperatures were largely overestimated previously, due to the inadequate treatment of both thermal and transverse fluctuations, and disorder. It 
appears that the Mn-Mn couplings can not explain the observed  RT-FM. We argue, that this can be attributed to  the interaction between large moments induced in the 
vicinity of the manganese. This  is similar to the non magnetic defect induced ferromagnetism reported in oxides, semiconductors and graphene/graphite.
\end{abstract}


\maketitle

\ The possibility to trigger and control ferromagnetism in transition metal doped oxides such as the widely studied ZnO\cite{fukumura,sato,kim,sharma,pereira}, 
TiO$_2$\cite{glaspell,errico}, HfO$_2$\cite{pucci,hong2005}, ZrO$_2$\cite{coey,hong,sangalli}, In$_2$O$_3$\cite{hongprb2006,phillip,peleckis} etc. has led to a 
huge development over these last 
years. These dielectric and transparent materials (large band gap) could unambiguously lead in the near future to a plethora of technological applications, in both areas 
of spintronics and 
opto-electronics, such as light emitting devices, magnetic sensors, detectors, ultra low-power memory devices etc. The interest in such materials has been growing 
tremendously during the last few years as demonstrated by the large amount of work found in the literature.
However, from the fundamental point of view, many questions are still to be clarified. The often reported room temperature ferromagnetism in these diluted magnetic 
oxides (DMO) is not always well understood. Recent experimental studies on Co doped ZnO\cite{jedrecy} interestingly revealed the existence of nano-sized 
ferromagnetic Co clusters, leading to critical temperatures of $\sim$300 K. The spinodal decomposition (alternating regions of low and high concentration of magnetic impurities) 
has been suggested to be the probable reason for the high Curie temperatures observed in some of these diluted materials\cite{katayama,bonanni,akash2012}. 
Nevertheless, the fact remains that the nature and origin of the ferromagnetism is often controversial and not always systematically reproducible in most of 
the DMO. From this point of view the case of ZrO$_2$, also known as synthetic diamond, is particularly interesting.

\ In a recent letter\cite{ostanin}, the authors have predicted that the cubic zirconia could exhibit relatively high Curie temperatures (beyond 500 K)  when doped with a relatively low 
concentration of Mn. It has also been shown that the ferromagnetism is robust against oxygen vacancies and defects. These interesting results have motivated many experimental 
studies. Unfortunately, many  have failed to observe ferromagnetism or could not reproduce the results from other groups. The ferromagnetism, when observed, 
appears to be metastable and sensitive to the growth and preparation conditions. Very often a non-ferromagnetic phase in Mn-doped ZrO$_2$ has been 
reported\cite{cravel,pucci,sandeep}. On the other hand, an inhomogeneous ferromagnetic phase has also been reported in zirconia thin films\cite{zippel}. The observed phase 
appeared to be independent of the Mn concentration and was attributed to intrinsic defects. In other words they have concluded that 
the Mn ions play a secondary role in the latter case, 
which contradicts the theoretical predictions. More recently, high Curie temperature ferromagnetism in cubic Mn-doped ZrO$_2$ thin films has been reported \cite{hong}. 
In this study, the authors have concluded that the ferromagnetism should result from the Mn-Mn exchange interactions. This in our view is not convincing since for a Mn concentration 
of about 5\% a huge moment of the order of 13.8 $\mu B$ per Mn was measured. In addition, in the monoclinic case room temperature ferromagnetism has also been observed 
in both doped and undoped host material. The conclusion was that in the latter case the FM is due to both the Mn-Mn couplings and the intrinsic defects. 
These defects were identified as structural defects, such as vacancies, located at the interface between film and substrate or at the surface. But the 
possibility of oxygen vacancies was ruled out from annealing measurements.
Ferromagnetism can indeed be induced by non-magnetic or intrinsic defects, this phenomenon is also known as $d^0$ ferromagnetism\cite{elfimov,osorio,venkatesan,esquinazi,georgesapl}. 
The non-trivial underlying physics involves many body aspects and still remains highly debated. In particular the competition between localization and electron-electron 
interaction which is at the 
origin of the magnetic moment formation in the vicinity of the defects. The induced moments could then eventually interact favorably, under certain conditions, and lead 
to long range ferromagnetic order.

\ In this letter, we propose to clarify  the role of Mn-Mn couplings in cubic Mn-doped ZrO$_2$, and show that this exchange mechanism can not explain the room temperature 
ferromagnetism observed. 
We demonstrate that the high Curie temperature predicted in a previous theoretical study\cite{ostanin} originates from an inaccurate treatment of both disorder 
(percolation effects not included) and thermal fluctuations (underestimated). For instance, we find that in the 5\% doped ZrO$_2$, the Curie temperature can not exceed 20 K. 
In addition, we present a detailed analysis of the nature of the magnetic excitations 
in defect free Mn doped zirconia. In the light of the current study, we will reconsider the interpretation of some existing experimental data. 

\begin{figure}[t]\centerline
{\includegraphics[width=3.2in,angle=-90]{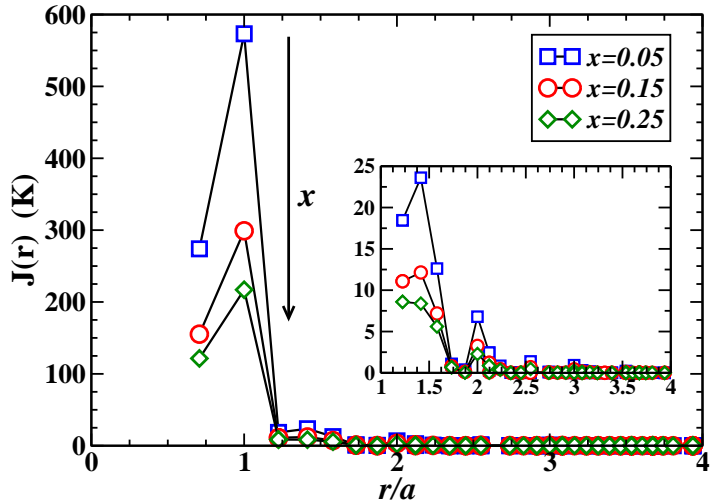}}
\caption{(Color online) Mn-Mn magnetic couplings in units of K in Mn doped ZrO$_2$ for various Mn concentrations $x$. The inset is a zoom of the extended 
exchange integrals \cite{kudrnovsky}.
}
\label{fig1} 
\end{figure} 

The Hamiltonian describing $N_{imp}$ interacting spins randomly distributed on a lattice of $N$ sites is given by the diluted random Heisenberg model,
\begin{eqnarray}
H_{Heis}=-\sum_{i,j} J_{ij}p_{i}p_{j} {\bf S}_{i}\cdot{\bf S}_{j}
\label{hamilt}
\end{eqnarray}
where the sum $ij$ runs over all sites and the random variable $p_i$ is 1 if the site is occupied by an impurity, otherwise it is 0. ${\bf S}_{i}$ is the 
localized Mn$^{2+}$ spin at site i.
 The couplings $J_{ij}$ depend on the Mn concentration\cite{kudrnovsky} and have been calculated from \textit{ab initio} method (Tight binding linear muffin tin 
 orbitals). These couplings were used in  Ref.\onlinecite{ostanin}. The above Hamiltonian (Eq.\ref{hamilt} ) is treated within the self-consistent local random 
 phase approximation (SC-LRPA)\cite{gbepl2005}. It is a semi-analytical approach based on finite temperature Green's functions. This approach has been discussed in details 
 and implemented successfully several times in the past (for more details see Ref.\onlinecite{satorev}). The agreement obtained with 
other approaches such as Monte Carlo simulations\cite{bergqvist2004,gb-manga} or site percolation statistics on random resistor networks\cite{akash2010} has established 
the reliability and accuracy of the SC-LRPA.

\begin{figure}[t]\centerline
{\includegraphics[width=3.40in,angle=-90]{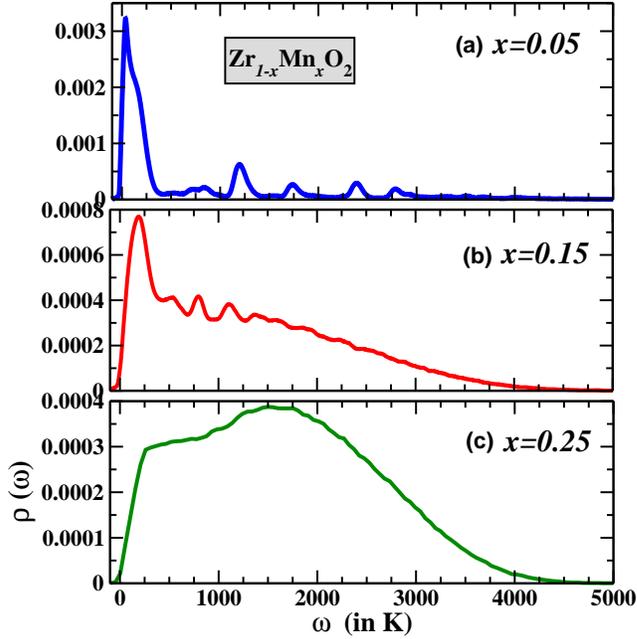}}
\caption{(Color online) Average magnon density of states (DOS) at $T$=0 K, corresponding to low, intermediate and large Mn concentrations: (a) $x$=$0.05$, 
(b) $x$=$0.15$ and (c) $x$=$0.25$. The energy ($x$ axis) is in unit of K.
}
\label{fig2}
\end{figure}

\ In Fig.\ref{fig1} , the Mn-Mn couplings for three different concentrations of Mn are plotted as a function of Mn-Mn distance. 
We observe that the couplings are all ferromagnetic but of very short range, beyond $r=2a$ they are almost negligible. Besides the first two couplings which are 
relatively large,  the others are much smaller. The short ranged nature of the couplings already suggests that the percolation effects are expected to play a 
crucial role. Note that these magnetic couplings were obtained within a standard Local Density Approximation (LDA) approach. Although it is well known that 
LDA often leads to incorrect results in both oxides and strongly correlated systems\cite{zunger}, we do not believe that the range of the couplings would drastically change 
with an improved treatment of the electronic correlations and thus affect qualitatively the nature of our conclusions.

\ In Fig. \ref{fig2}  we show the magnon density of states (DOS) $\rho(\omega)$ for three different concentrations of Mn: low, intermediate and relatively large 
concentration. First, we observe that the overall shape  changes drastically as we increase the Mn concentration. For the lowest concentration one sees that 
the magnon DOS  has a significant weight only in the relatively low energy region. In addition 
it exhibits a multiple peak structure separated by ``pseudo gap'' regions (very low density of states). These peaks result from weakly coupled clusters of Mn. Indeed, 
as seen previously both the nearest and next-nearest neighbor couplings are much larger than the others. For $x=0.15$ the magnon DOS has already changed 
significantly. Besides a relatively narrow peak at low energy the magnon DOS now has a significant weight which extends to relatively high energies. We still 
observe some peak structure for the intermediate energies. For the largest concentration $x=0.25$, the magnon DOS exhibits now a broad peak at 1500 K and a 
flat region between 250 K to 1000 K. The significant change in the DOS when the Mn concentration varies from $x=0.15$ to $x=0.25$, is a signature of the very short 
ranged nature of the exchange interactions.

\begin{figure}[t]\centerline
{\includegraphics[width=3.2in,angle=-90]{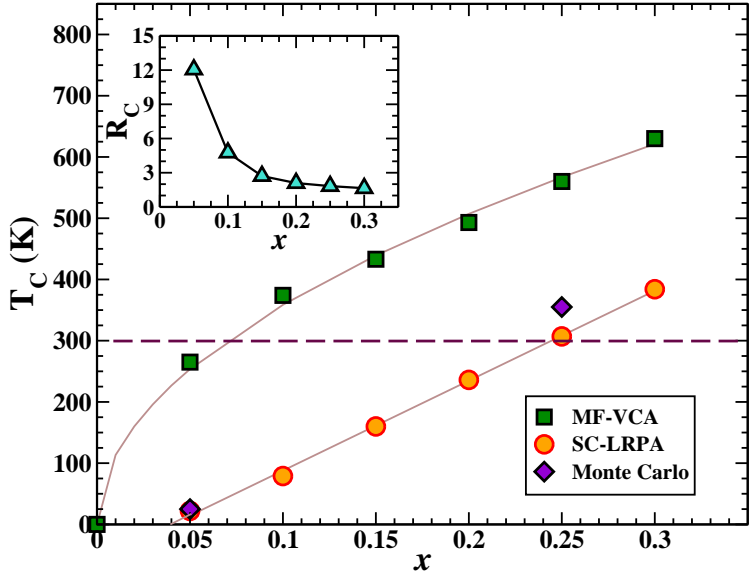}}
\caption{(Color online) Curie temperature as a function of the Mn concentration $x$, calculated within  (i) Mean Field VCA approximation (see text), 
(ii) SC-LRPA and (iii) Monte Carlo\cite{bergqvist}. For SC-LRPA calculations the size of the symbols provides the error bar. The continuous lines are a guide 
to the eye. The inset is the ratio R$_C$=$\frac{T_C^{\mathsf {VCA}}}{T_C^{\mathsf {LRPA}}}$.
}
\label{fig3}
\end{figure} 

\begin{figure}[t]\centerline
{\includegraphics[width=3.9in,angle=-90]{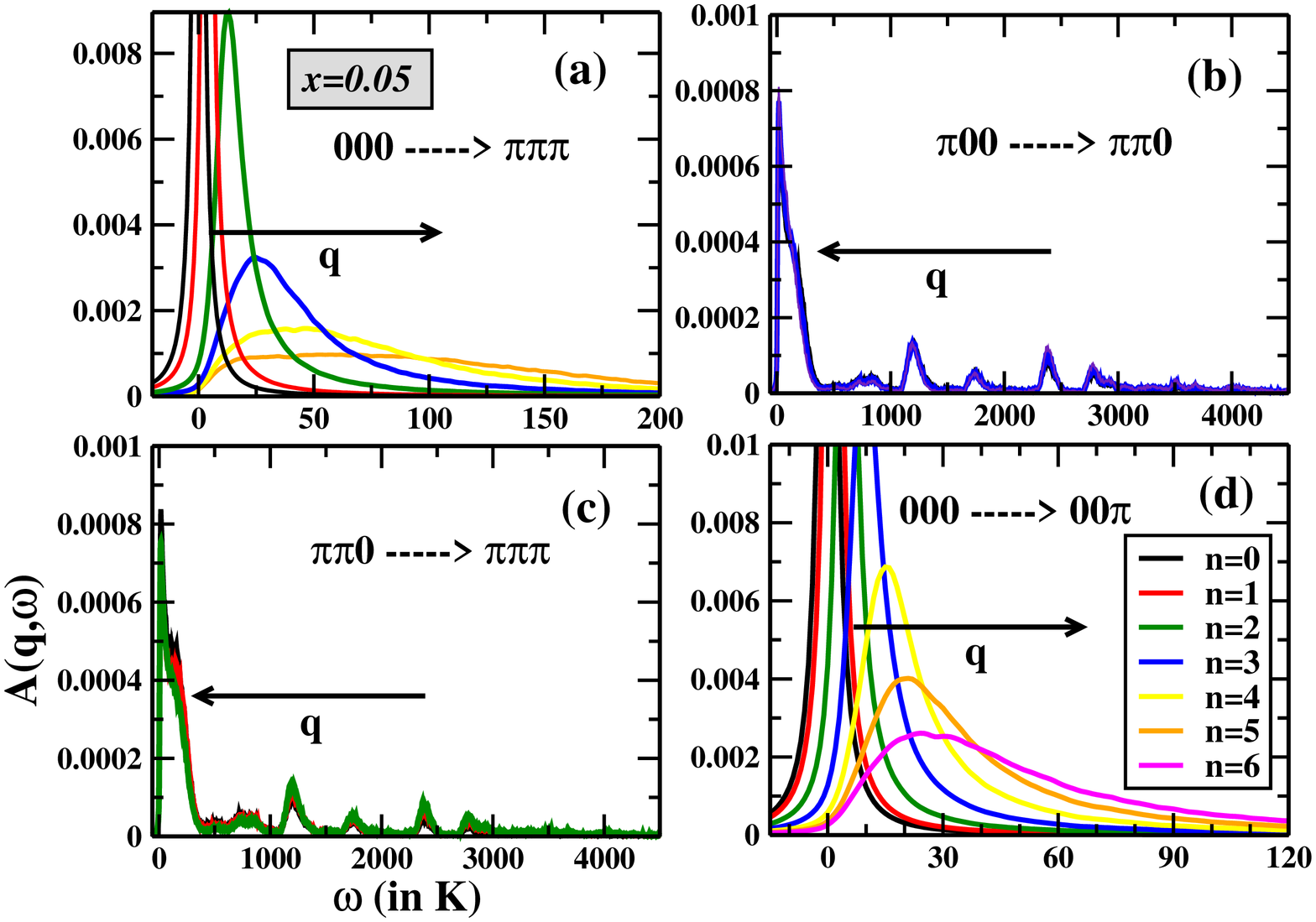}}
\caption{(Color online) Spectral function averaged over the disorder A($\bf q$,$\omega$) for different directions of  the Brillouin zone. In each case, we show 
only few values of  $\bf q$. In (a) and (d) only the lowest $\bf q$ values are shown starting from the $\Gamma$ point.  In (d) the momentum correspond to 
$\bf q$=$(0,0,\frac{2\pi n}{La})$ where n=0,1,2,..6. The Mn concentration is set to $x$=0.05. The system contains $N=4L^3$ sites where $L$=32.}
\label{fig4}
\end{figure}

\ Let us now discuss the effects of a proper treatment of 
both disorder and thermal/transverse fluctuations on the Curie temperatures, which is depicted in Fig.\ref{fig3}. We compare the self-consistently calculated $T_C$ 
to that given in Ref.\onlinecite{ostanin} which was obtained within 
the mean-field virtual crystal approximation (VCA). Its expression reads $T_C^{VCA}=\frac{2}{3} x\sum_{i} z_{i} J_{0i}$, where the sum runs over 
the shells and $z_{i}$ denotes the number of sites on the $i$-th shell. First we observe that the critical temperatures obtained within SC-LRPA are much smaller. In fact  
for 5\% of Mn, $T_C^{VCA}$ is 1200\% larger than the true critical temperature,  it is about 500\% larger for 10\% of Mn and it is 
two times larger even for 20\% of Mn. Note that the SC-LRPA $T_C$ is also in very good agreement with the Monte Carlo calculations\cite{bergqvist}.
While, within mean field, $T_C$ beyond room 
temperature can be already achieved for only 5\% of Mn, when disorder and thermal fluctuations are properly included room temperature ferromagnetism 
is only possible beyond 25\% of Mn. The artificially  high critical temperatures obtained within mean field is a consequence of the large values of the nearest and 
next nearest neighbor couplings, depicted in Fig.\ref{fig1}. However, when the percolation effects are taken into account these couplings are expected to play only a 
negligible role in the dilute regime\cite{richard2010}. Note that $T_C$ is well described by $A(x-x_c)$. This is in contrast to what was found for Mn-doped GaAs, where 
the best fit was of the form $A'(x-x_c)^{1/2}$ (Ref.\onlinecite{georges2005}). We believe that the origin of the linear like behavior in the present case arises from the strong dependence 
of the relevant couplings, for $r/a\ge1$, on $x$ (see Fig.1); unlike in the case of Mn-doped GaAs, where the typical couplings which controls the critical temperature were 
found to vary weakly with $x$.

\ These results clearly show that the room temperature ferromagnetism observed in about 5\% Mn doped cubic ZrO$_2$ reported in Ref.\onlinecite{hong} can not result 
from Mn-Mn couplings, in case of homogeneous distribution of dopants. The Mn-Mn exchange couplings mechanism would only lead to a critical temperature of about 
20 K. We believe that the room temperature ferromagnetism 
observed in cubic Mn-doped zirconia could be of $d^0$ nature\cite{elfimov,osorio,venkatesan,esquinazi,georgesapl}. The localized spin of Mn$^{2+}$ plays 
an irrelevant role. The substitution of Zr$^{4+}$ by Mn$^{2+}$ would lead to the formation of large magnetic moments around the Mn-atoms, 
interacting with each other ferromagnetically via relatively extended couplings. This scenario would be consistent with the large moment of 13.8 $\mu B$ 
measured in Ref.\onlinecite{hong}. Another possibility for the room temperature phenomenon observed in this case might be the presence of spinodal 
decomposed or inhomogeneous phases. However, a more careful and detailed structural analysis of the samples is necessary to confirm this scenario.

\begin{figure}[t]\centerline
{\includegraphics[width=3.80in,angle=-90]{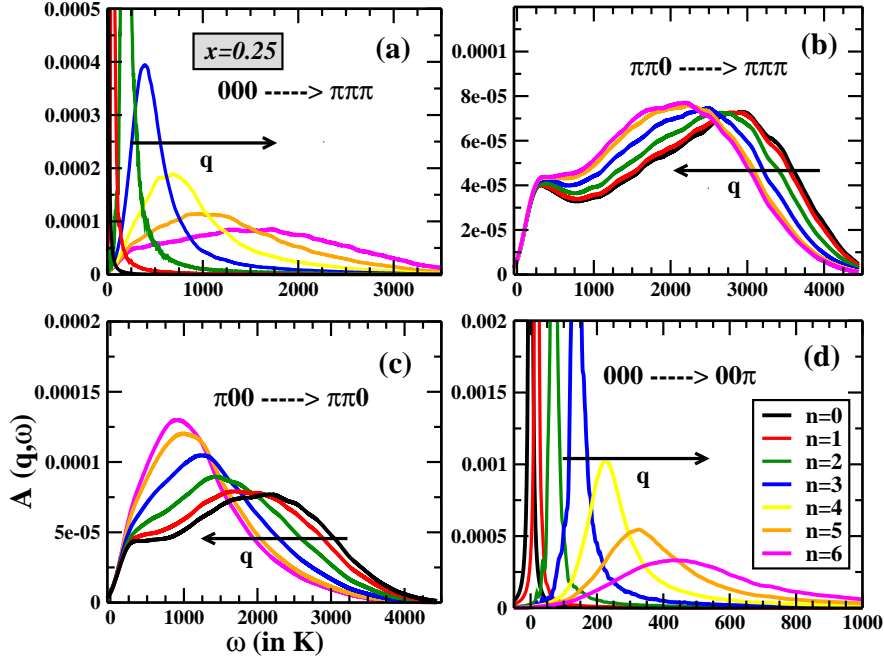}}
\caption{(Color online) 
As in the previous figure, spectral function averaged over the disorder $A(\bf q$,$\omega$) for different directions of  the Brillouin zone. The Mn concentration is 
set to $x$=0.25. The system contains $N=4L^3$ sites where $L$=20.}
\label{fig5}
\end{figure}

\ We now analyze the nature of the magnon excitations, with a particular focus on the long wavelength limit. In 
Fig.\ref{fig4} we have plotted, for a fixed concentration of 5\%  of Mn and at $T$=0 K the spectral function averaged over disorder. It is given by the following 
expression ${A}({\bf q},\omega) =-\left\langle \frac{1}{2\pi S} {\rm Im} G({\bf q},\omega)\right\rangle_c$, where $G({\bf q},\omega)$ is the Fourier transform of the retarded 
Green's function $G_{ij}(t)=-i\theta(t)\langle[S_i^+(t),S_j^-(0)]\rangle$, ($\langle ...\rangle$ denotes the thermal average and $\langle ...\rangle_c$ the average 
over disorder configurations). Fig.\ref{fig4}(a)-(d) represent $A(\bf q$,$\omega$) in various directions in the Brillouin zone (BZ). We observe in (a) and (d) that 
well-defined magnons are visible only in a very 
narrow region around the $\Gamma$ point. The width of the excitations is found to 
increase strongly as we move away from the $\Gamma$ point. In (b) and (c) we observe that 
 $A(\bf q$,$\omega$) is insensitive to the momentum. For these momenta, $A(\bf q$,$\omega$) is identical to the density of states shown in Fig.\ref{fig2}. 
The magnon excitations are completely incoherent in these regions of the BZ. This behavior also indicates that these high energy modes (peaks) are completely localized. 
In Fig.\ref{fig5}, the disorder averaged spectral function is now plotted for a larger concentration of Mn, $x=0.25$. Qualitatively this figure resembles the previous one, 
besides the fact that the long wavelength modes are now well defined for a broader range of momentum q (the system size was $L$=32 in Fig.\ref{fig4} and $L$=20 in the 
present case). We will come back
 to this point in a more quantitative way in what follows. In both (b) and (c), and in contrast to the low concentration case, we now observe a clear strong q dependence of 
 $A(\bf q$,$\omega$), more pronounced in (c). This indicates that the nature of the high energy modes differs from those seen in the low concentration case. However, 
 even for this reasonably large concentration no well defined magnons exist in these regions of the BZ.

\begin{figure}[t]\centerline
{\includegraphics[width=3.5in,angle=-90]{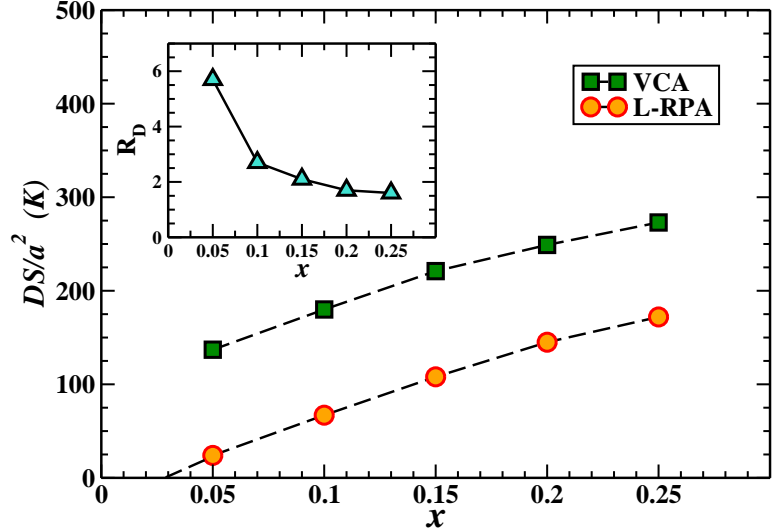}}
\caption{(Color online) 
Averaged spin stiffness $D$ as a function of the Mn concentration $x$ in Mn doped ZrO$_2$. The calculations were performed both within VCA and SC-LRPA. The inset
 is the ratio $R_D$=$\frac{D^{\mathsf{VCA}}}{D^{\mathsf{LRPA}}}$. $a$ is the lattice parameter.
}
\label{fig6}
\end{figure}
\begin{figure}[t]\centerline
{\includegraphics[width=3.5in,angle=-90]{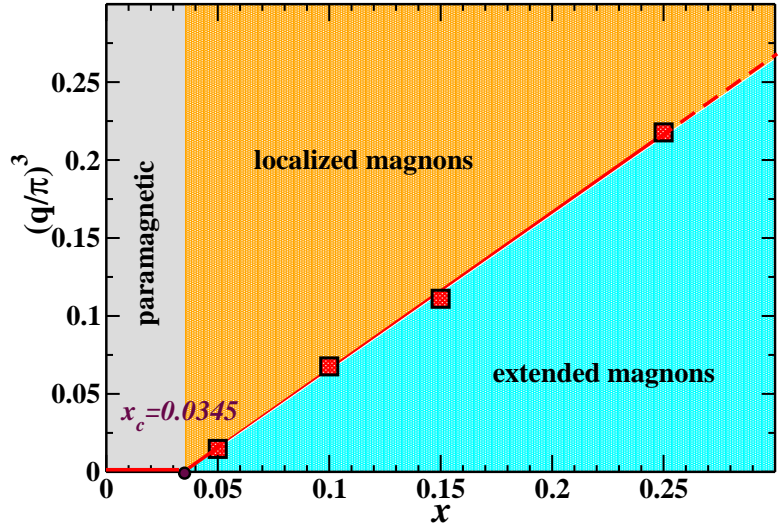}}
\caption{(Color online) 
Phase diagram for the low energy magnon excitations.The symbols separate the extended well-defined magnon modes from the localized excitations (see text). $x_c$ denotes 
the percolation threshold which is approximately 0.0345. Here we consider $\bf q$ in the $(1,0,0)$ direction.}
\label{fig7}
\end{figure}

 

\ In order to get a better insight of the low energy excitations we have plotted in Fig.\ref{fig6} the spin stiffness $D$ as a function of the Mn concentration. 
We remind that in the long wavelength limit the magnon energy scales like $\omega (${\bf q}$)$=$Dq^2$. To evaluate quantitatively the effects of percolation and thermal 
fluctuations, both the mean field value and that calculated from the finite size analysis of $A(\bf q$,$\omega$) are shown. Note that the mean-field value of the spin stiffness 
is $D^{VCA}=\frac{1}{3S} x\sum_{i} z_{i} r_{i}^2J_{0i}$, where the sum runs over the shells and $r_i$ is the distance between the sites of the i-th shell and a site at the 
origin. We observe a dramatic effect on $D$ for the low concentrations. Indeed, for $x=0.05$ the mean field spin stiffness is six times larger 
than that obtained within the local RPA. Even for the 10\% doped system it is almost three times larger, and twice in the 20\% case. 
We also observe that the extrapolation of the spin stiffness vanishes at approximately $x \approx 0.03$ (percolation threshold).
This agrees very well with the value that could be estimated from the critical temperature (Fig.\ref{fig3}). We expect indeed, both $T_C$ and $D$ to vanish 
exactly at the percolation threshold.

\ Finally we provide a quantitative estimate of the region of the BZ where well defined magnons exist. We focus only 
in the (1,0,0) direction here. One way to extract accurately the critical line that separates extended from localized magnons would be to calculate the 
typical density of states\cite{dobrosav1,dobrosav2}. This method allows to determine the mobility edge for the metal-insulator transition in the Anderson model. 
As this approach is more demanding, for the sake of simplicity, we consider the following alternative natural criterion: 
the magnon modes are well defined only if their linewidth $\gamma(${\bf q}$)$ (inverse of the lifetime) is smaller than their energy $\omega (${\bf q}$)$.
 The results are depicted in Fig.\ref{fig7}. We find that the line that separates the well defined magnons from localized magnon modes is given by
$q_{c}=A(x-x_{c})^{1/3}$, where $x_{c}$ is found to be 0.0345. This is in excellent agreement with that extracted from the spin stiffness and the critical 
temperature. One immediately sees, for the 25\% doped case the volume of well defined magnons is fifteen times larger than that of the 5\% doped zirconia.

To conclude, we have shed light on some existing controversies on the origin of room temperature ferromagnetism in Mn doped zirconia. We have demonstrated that the 
Mn-Mn interactions alone can not give rise to room temperature ferromagnetism in ZrO$_2$, in contrast to what was predicted 
in previous theoretical studies. In particular, it has been shown that the 5\% Mn  doped systems is close to percolation and its critical temperature can not exceed 20 K. 
Room temperature  ferromagnetism is reachable only for large Mn concentration, beyond 25\%. Thus, in our view, the room temperature ferromagnetism observed experimentally in 
cubic Mn-doped zirconia could be of $d^0$ nature, and the spin of Mn plays only a secondary role. The question is whether the manganese induces large magnetic moment in its surrounding or the ferromagnetism is due to other intrinsic defects. 
However, we do not rule out the possibility of nanoscale spinodal decomposition which can also lead to 
room temperature ferromagnetism, even in the 5\% doped case, for effective short-ranged interactions. It will be interesting to investigate these issues experimentally. 
In addition, we have presented a detailed analysis of the energy and 
linewidth of the spin excitations with a particular focus on the long wavelength limit (spin stiffness).  It is found that for 5\% doped zirconia extended magnons exist only 
within about 2\% of the volume of the Brillouin zone, whilst it reaches approximately 25\% for 25\% doped zirconia. We believe that our study should also be relevant for other 
compounds, namely transition metal ion doped oxides such as TiO$_2$, HfO$_2$, SnO$_2$ or In$_2$O$_3$.

\acknowledgments
This work was supported by the EU within FP7-PEOPLE-ITN-2008, Grant number 234970 Nanoelectronics: Concepts, Theory and Modelling. We thank J. Kudrnovsk\'y for 
providing us with the Mn-Mn couplings in ZrO$_2$. We would also like to thank L. Bergqvist for the Monte Carlo data.


\begin{thebibliography}{99}
\bibitem{fukumura} T. Fukumura et. al. Appl. Phys. Lett. \textbf{75,} 3366, (1999).
\bibitem{sato} K. Sato and K. Yoshida, Jpn. J. Appl. phys. \textbf{40,} L334 (2001).
\bibitem{kim} J. H. Kim, J. Appl. Phys. \textbf{92,} 6066 (2002).
\bibitem{sharma} P. Sharma et al. Nat. Mat. \textbf{2,} 673 (2003).
\bibitem{pereira} L. M. C. Pereira et al. J. Appl. Phys. \textbf{113,} 023903 (2013).
\bibitem{glaspell} G. Glaspell, A. B. Panda, and M. S. El-Shall, J. Appl. Phys. \textbf{100,} 124307 (2006). 
\bibitem{errico} L. A. Errico, M. Renteria, and M. Weissmann, Phys. Rev. B \textbf{72,} 184425 (2005).
\bibitem{pucci} A. Pucci, G. Clavel, M-G. Willinger, D. Zitoun, and N. Pinna, J. Phys. Chem. C \textbf{113,} 12048 (2009).
\bibitem{hong2005} N. H. Hong, J. Sakai, N. Poirot, and A. Ruyter, Appl. Phys. Lett. \textbf{86,} 242505 (2005).
\bibitem{coey} J. M. D. Coey, M. Venkatesan, P. Stamenov, C. B. Fitzgerald, L. S. Dorneless, Phys. Rev. B \textbf{72,}  024450 (2005).
\bibitem{hong} N. H. Hong, C.-K. Park, A. T. Raghavender, O. Ciftja, N. S. Bingham, M. H. Phan, and H. Srikanth, J. Appl. Phys. \textbf{111,} 07C302 (2012).
\bibitem{sangalli} D. Sangalli et al. Eur. Phys. J. B \textbf{86,} 211 (2013).
\bibitem{hongprb2006} N. H. Hong et al. Phys. Rev. B \textbf{73,} 132404 (2006).
\bibitem{phillip}  J. Phillip et al.  Appl. Phys. Lett. \textbf{85,} 777 (2004). 
\bibitem{peleckis} G. Peleckis, X. Wang, and X. D. Shi, Appl. Phys. Lett. \textbf{89,} 022501 (2006).
\bibitem{jedrecy} N. Jedrecy, H. J. von Bardeleben, and D. Demaille, Phys. Rev. B \textbf{80,} 205204 (2009).
\bibitem{katayama} H. Katayama-Yoshida et. al. Phys. Stat. Sol. (a) \textbf{204,} 15 (2007).
\bibitem{bonanni} A. Bonanni and T. Dietl, Chem. Soc. Rev. \textbf{39,} 528 (2010).
\bibitem{akash2012} A. Chakraborty, R. Bouzerar, S. Kettemann, and G. Bouzerar, Phys. Rev. B \textbf{85,} 014201 (2012).
\bibitem{ostanin}  S. Ostanin, A. Ernst, L. M. Sandratskii, P. Bruno, M. Dane, I. D.
Hughes, J. B. Staunton, W. Hergert, I. Mertig, and J. Kudrnovsk\'y, Phys. Rev. Lett. \textbf{98,} 016101 (2007).
\bibitem{cravel} G. Cravel, M.-G. Willinger, D. Zitoun, and N. Pinna, Eur. J. Inorg. Chem. 2008, \textbf{863,} (2008).
\bibitem{sandeep} S.K.  Srivastava, P. Lejay, B. Barbara, O. Boisron, S. Pailhes and G. Bouzerar, J. Appl. Phys. \textbf{110,} 043929 (2011).
\bibitem{elfimov} I.S. Elfimov, S. Yunoki and G.A. Sawatzky, Phys. Rev. Lett. \textbf{89,} 216403 (2002). 
\bibitem{osorio} J. Osorio-Guillen, S. Lany, S. V. Barabash, A Zunger, Phys. Rev. Lett \textbf{96,} 107203, 2006.
\bibitem{venkatesan} M. Venkatesan et al. Nature (London) \textbf{430,} 630 (2004), J.M.D. Coey et al., Phys. Rev. B, \textbf{72,} 024450 (2005).
\bibitem{esquinazi} P. Esquinazi et al. Phys. Rev. Lett., \textbf{87,} 227201 (2003).
\bibitem{georgesapl} F. Maca, J. Kudrnovsk\'y, V. Drchal, and G. Bouzerar, Appl. Phys. Lett. \textbf{92,} 212503 (2008); G. Bouzerar and T. Ziman, 
Phys. Rev. Lett. \textbf{96,} 207602 (2006).
\bibitem{zippel} J. Zippel, M. Lorenz, A. Setzer, G. Wagner, N. Sobolev, P. Esquinazi, and M. Grundmann, Phys. Rev. B \textbf{82,}125209 (2010).
\bibitem{kudrnovsky} J. Kudrnovsk\'y, \textit{private communication}. 
\bibitem{gbepl2005} G. Bouzerar, T. Ziman, and J. Kudrnovsk\'y, Europhys. Lett. \textbf{69,} 812 (2005).
\bibitem{satorev} K. Sato et. al., Rev. Mod. Phys. \textbf{82,} 1633 (2010).
\bibitem{bergqvist2004} L. Bergqvist, O. Eriksson, J. Kudrnovsk\'y, V. Drchal, P. Korzhavyi and I. Turek, Phys. Rev. Lett. \textbf{93,} 137202 (2004).
\bibitem{gb-manga} G. Bouzerar and O. Cepas, Phys. Rev. B \textbf{76,} 020401R (2007).
\bibitem{akash2010} A. Chakraborty and G. Bouzerar, Phys. Rev. B \textbf{81,} 172406 (2010). 
\bibitem{zunger} see for example A. Zunger, S. Lany, and H. Raebiger, Physics \textbf{3,} 53 (2010). 
\bibitem{bergqvist} L. Bergqvist, \textit{private communication}. 
\bibitem{richard2010} R. Bouzerar and G. Bouzerar, New J. of Physics \textbf{12,} 053042 (2010). 
\bibitem{georges2005} G. Bouzerar, T. Ziman, and J.  Kudrnovsk\'y, Phys. Rev. B \textbf{72,} 125207 (2005).
\bibitem{dobrosav1} V. Dobrosavljevi\'c, A. A. Pastor, and B. K. Nikolic, Europhys. Lett. \textbf{62,} 76 (2003).
\bibitem{dobrosav2} V. Dobrosavljevi\'c and G. Kotliar, Phys. Rev. Lett. \textbf{78,} 3943 (1997).

\end{thebibliography}
\end{document}